\documentclass[%
 reprint,
 amsmath,amssymb,
 aps,
prl,
floatfix,
]{revtex4-1}

\usepackage{graphicx}% Include figure files
\usepackage{dcolumn}% Align table columns on decimal point
\usepackage{bm}% bold math
\begin{document}

\preprint{APS/123-QED}

\title{Path Integral Molecular Dynamics for Bosons}%:\\with Forced Linebreak}% Force line breaks with \\
%\thanks{A footnote to the article title}%

\author{Barak Hirshberg}
 \affiliation{Department of Chemistry and Applied Biosciences, ETH Zurich, 8092 Zurich, Switzerland}%Lines break automatically or can be forced with \\
\affiliation{Institute of Computational Sciences, Universit\`{a} della Svizzera italiana, via G. Buffi 13, 6900 Lugano, Switzerland}%

\author{Valerio Rizzi}
 \affiliation{Department of Chemistry and Applied Biosciences, ETH Zurich, 8092 Zurich, Switzerland}%Lines break automatically or can be forced with \\
\affiliation{Institute of Computational Sciences, Universit\`{a} della Svizzera italiana, via G. Buffi 13, 6900 Lugano, Switzerland}%

\author{Michele Parrinello}
 \email{parrinello@phys.chem.ethz.ch}
 \affiliation{Department of Chemistry and Applied Biosciences, ETH Zurich, 8092 Zurich, Switzerland}%Lines break automatically or can be forced with \\
\affiliation{Institute of Computational Sciences, Universit\`{a} della Svizzera italiana, via G. Buffi 13, 6900 Lugano, Switzerland}%
\affiliation{Italian Institute of Technology, Via Morego 30, 16163 Genova, Italy}

\date{\today}% It is always \today, today,
             %  but any date may be explicitly specified

\begin{abstract}
Trapped Bosons exhibit fundamental physical phenomena and are potentially useful for quantum technologies. We present a method for simulating Bosons using path integral molecular dynamics. A main challenge for simulations is including all permutations due to exchange symmetry. We show that evaluation of the potential can be done recursively, avoiding explicit enumeration of permutations, and scales cubically with system size. The method is applied to Bosons in a 2D trap and agrees with essentially exact results. An analysis of the role of exchange with decreasing temperature is also presented.
\end{abstract}

%\keywords{Suggested keywords}%Use showkeys class option if keyword
                              %display desired
\maketitle

%\section{Introduction}
Trapped cold atoms are fascinating systems that allow studying fundamental physical phenomena. These range from the observations of Bose-Einstein condensation in macroscopic systems~\cite{Anderson1995,Davis1995} to the study of exotic supersolids~\cite{Tanzi2019} and the formation of ultra-cold diatomic molecules~\cite{Regal2003, Ni2008}. %and their chemical transformations~\cite{Bohn2017}. Such systems are also at the heart of emerging quantum technologies~\cite{Nshii2013}.
Experiments can control many properties of such systems. They can be studied in optical lattices with single-atom resolution~\cite{Bakr2010,Sherson2010}, at varying interaction strengths~\cite{Paredes2004,Kinoshita2004,Zurn2012}, under the influence of external fields~\cite{Buchler2007} and more. Developing numerical simulations with microscopic resolution of such systems is thus highly desirable. In this Letter, we report a new method for performing simulations of Bosons using path integral molecular dynamics (PIMD)~\cite{Parrinello1984}. 

Symmetry under permutation of indistinguishable particles is a fundamental property of many-body quantum systems. Therefore, methods which include exchange effects are an important goal in physics and chemistry. The path integral formulation of quantum mechanics~\cite{Feynman2005} is a powerful tool for studying many-body systems~\cite{Markland2018}. Applications of path integral methods often neglect exchange effects, treating the particles as distinguishable. This approximation leads to the well-known isomorphism between the partition function of a quantum-mechanical system and that of a classical system, in which each particle is represented by a ring-polymer composed of $P$ beads connected through harmonic springs~\cite{Chandler1981}. %Beads representing different particles interact through an interaction potential, scaled by the number of beads. 
The partition function of the classical system can be sampled using molecular dynamics (MD)~\cite{Parrinello1984} or Monte Carlo (MC)~\cite{Pollock1984}. 

\begin{figure}
\includegraphics[width=1.0\columnwidth]{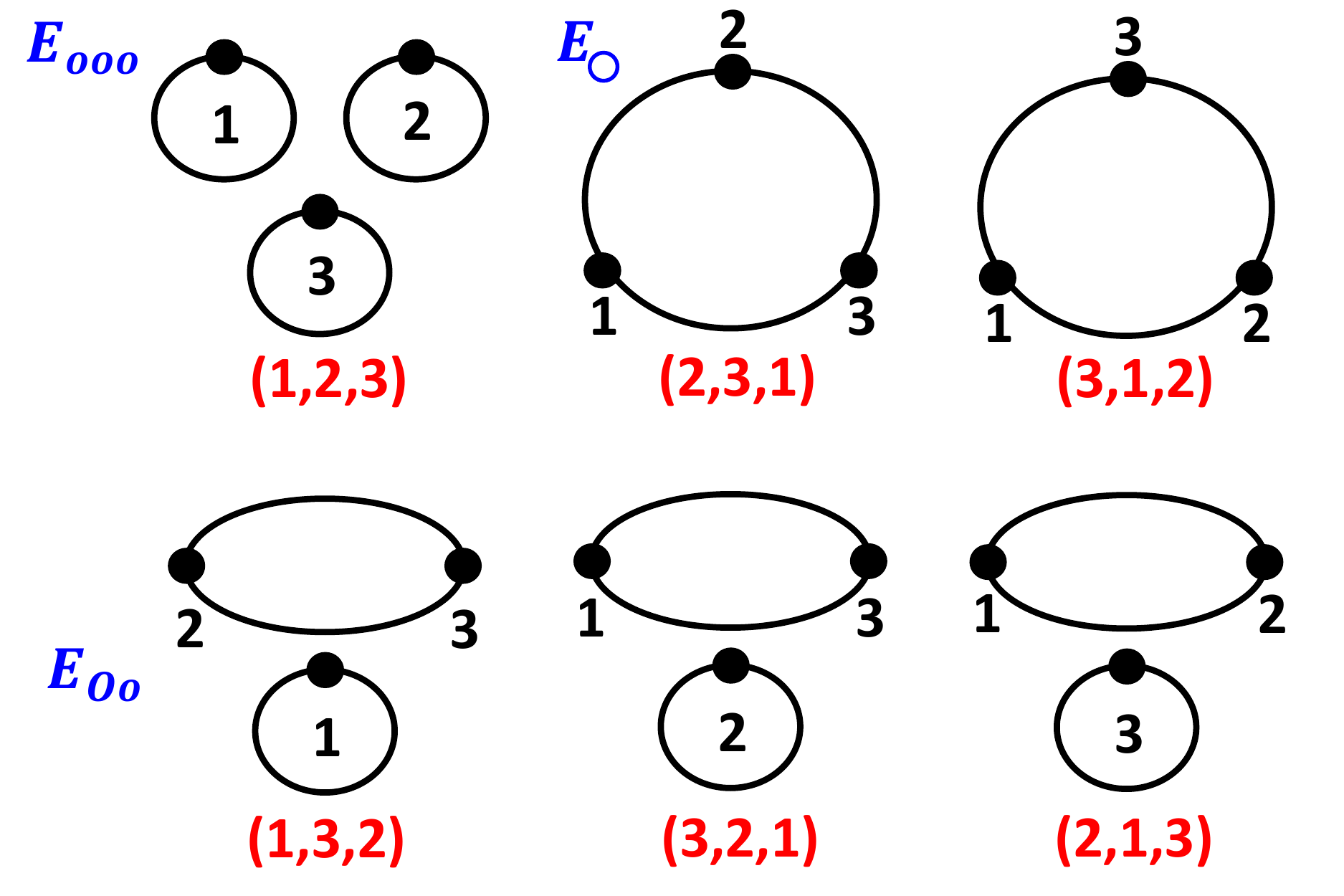}% Here is how to import EPS art
\caption{\label{fig:Fig1} All permutations for $N=3$ and their corresponding ring polymer configurations (see Equation~\ref{eq:VB3_wdiagrams}). }
\end{figure}

In the case of indistinguishable particles, one must include all possible $N!$ permutations ~\cite{Ceperley1995}. This leads to ring-polymer configurations in which permuted particles are connected sequentially into longer rings (see Figure~\ref{fig:Fig1}). For Bosons the weights of all configurations are positive while for Fermions they can be negative, giving rise to the infamous sign problem~\cite{DuBois2014, Runeson2018}. Here we focus on Bosons, the treatment of Fermions being beyond the scope of this Letter.
Enumerating all Bosonic ring-polymer configurations is impractical in all but the smallest systems. To alleviate this difficulty, one can address it as a sampling problem and develop Monte Carlo algorithms that introduce permutations. The most notable examples are the pioneering applications of path integral Monte Carlo (PIMC) to superfluid Helium~\cite{Ceperley1995}. Significant advancement in efficiency was achieved with the introduction of the worm algorithm~\cite{Boninsegni2006}. 

Far less progress has been made in the use of MD to sample the Bosonic partition function. Miura and Okazaki have suggested a method for performing PIMD simulations of Bosons~\cite{Miura2000}. However, their approach involved calculating a permanent. Since this is equivalent to enumerating all possible permutations, the method was only applied to very small systems. In this Letter, we show how to evaluate the potential energy and forces acting on the particles avoiding calculating the permanent. This results in an algorithm that scales cubically with system size, allowing larger systems to be studied. The resulting approach is simpler than available rather-complex sampling algorithms. Developing alternatives to MC sampling can extend the applicability of path integral methods. 

In the following, we first present the method and benchmark it against analytical results for up to 64 non-interacting Bosons and numerical diagonalization of the Hamiltonian for small interacting systems~\cite{Mujal2017}. We also apply the method to 32 interacting particles in a 2D trap, for which the number of permutations to be considered is already prohibitive. Finally, we present an approach to analyze the importance of exchange effects from the different ring-polymer configurations.

%\section{Method}

We consider a system of $N$ identical particles of mass $m$ at inverse temperature $\beta = (k_B T)^{-1} $. The interaction  between the particles is denoted as $V(\bm r_1,...,\bm r_N)$. 

% For simplicity, we present the equations for non-interacting particles but the extension for interacting systems is given in the SI. 

Neglecting effects due to quantum symmetry, the path integral representation of the partition function is given~\cite{tuckerman2010} by 

% \begin{widetext}
\begin{equation}
% Z_D^{(N=3)} \sim \int e^{-\beta V_{ooo}} dR_1 dR_2 dR_3  ,
Z_D^{(N)} \sim \int e^{-\beta \left( V_D^{(N)} + U \right)} dR_1 ... dR_N  ,
\label{eq:ZD}
\end{equation}
% \end{widetext}
% Note the open one in Eq.~(\ref{eq:one}).
where $R_l$ represents collectively the coordinates $(\bm r_l^{1},...,\bm r_l^{P})$ of the $P$ beads composing the ring polymer corresponding to particle $l$ and $dR_l=d\bm r_l^1...d\bm r_l^P$. The exact result is obtained in the limit $P \rightarrow \infty $.
% $ V_{ooo}(R_1,R_2,R_3) $ is the sum of spring energies over all ring polymer representing different particles, given by
In Equation~\ref{eq:ZD}, the interaction potential between beads $j$ of different particles is scaled by the number of beads and $U(R_1,...,R_N) = \frac{1}{P} \sum_{j=1}^P V(\bm r_1^{j},...,\bm r_N^{j})$.
The potential $ V_D^{(N)}(R_1,...,R_N) $ is the sum of ring polymer spring energies $E_o(R_l)$ for each particle,

% \begin{widetext}
\begin{align}
% V_{ooo}(R_1,R_2,R_3) = \sum_{l=1}^3 V_o(R_l)  \nonumber\\ 
% = \frac{1}{2}m \omega_P^2 \sum_{l=1}^3 \sum_{j=1}^P \left( \bm r_l^{j+1} - \bm r_l^{j} \right) ^2.
& V_D^{(N)}(R_1,...,R_N) = \sum_{l=1}^N E_o(R_l)  \nonumber\\ 
& \equiv \frac{1}{2}m \omega_P^2 \sum_{l=1}^N \sum_{j=1}^P \left( \bm r_l^{j+1} - \bm r_l^{j} \right) ^2.
\label{eq:VD}
\end{align}
% \end{widetext}
% Note the open one in Eq.~(\ref{eq:one}).
The bead index in Equation~\ref{eq:VD} is cyclic, $ \bm r_l^{P+1} = \bm r_l^1 $, and $ \omega_P = \sqrt{P}/\beta\hbar $ is the frequency of the springs connecting the beads in the ring polymer.

To include exchange effects one must consider all permutations. This results in a sum over $N!$ terms in the partition function~\cite{Miura2000}. Each term in the sum corresponds to a configuration of the ring polymers in which some of the particles are connected into longer rings~\cite{Runeson2018}. Figure~\ref{fig:Fig1} shows all configurations for $N=3$. However, when integrated over all particles, some of the configurations contribute equally to the partition function. Therefore, one can sum over a much smaller number of terms arising from the number of unique configurations for $N$ particles~\cite{Lyubartsev1993}. Each term is weighted by the number of different permutations which lead to the same configuration. This observation has been used recently in attempts to overcome the Fermion sign problem in PIMC simulations~\cite{Voznesenskiy2009, DuBois2014}. For Bosons, all configurations give a positive contribution and no such sign problem arises.

% Miura and Okazaki have shown\cite{Miura2000} that this leads to the following partition function for $N$ Bosons,

% % \begin{widetext}
% \begin{equation}
% Z_B^{(N)} \sim \int e^{-\beta V_D^{(N)}} perm(\tilde{A}) dR_1...dR_N ,
% \label{eq:ZB3}
% \end{equation}
% % \end{widetext}
% % Note the open one in Eq.~(\ref{eq:one}).

% where $\tilde{A} $ is an $N$x$N$ given by $ \tilde{A}_{ij}=A_{ij}/A_{ii}$ and

% % \begin{widetext}
% \begin{equation}
% A_{ij}=e^ {-\frac{\beta}{2} m \omega_P^2 \left( \bm r_i^{P} - \bm r_j^{1} \right)^2 }.
% \label{eq:Amatrix}
% \end{equation}
% % \end{widetext}
% % Note the open one in Eq.~(\ref{eq:one}).

% Since Equation~\ref{eq:ZB3} involves the calculation of a permanent, this approach for performing PIMD simulations for Bosons has only been applied to very small systems so far. In the context of PIMC simulations of fermions it was shown \cite{Lyubartsev1993, Voznesenskiy2009} that
% Equation~\ref{eq:ZB3} can be significantly simplified by summing only over a much smaller number of non-equivalent configurations, belonging to the symmetric group for $N$ particles. 

For example, the partition function for $N=3$ Bosons can be written~\cite{tuckerman2010} as

% \begin{widetext}
\begin{align}
Z_B^{(3)} & \sim \int e^{-\beta \left (V_B^{(N=3)}+ U \right)} dR_1...dR_3 ,
\label{eq:ZB3}
\end{align}
% \end{widetext}
% Note the open one in Eq.~(\ref{eq:one}).
% where the different configurations a
% \begin{align}
% & E_{ooo} \equiv V_D^{(3)} \nonumber \\
% & E_{Oo} \equiv E_o(R_1)+ E^{(2)}(R_2,R_3) + U(R_1,R_2,R_3) \nonumber \\
% & E_{\bigcirc} \equiv E^{(3)}(R_1,R_2,R_3) + U(R_1,R_2,R_3), 
% \end{align}
where the potential $V_B^{(N=3)}(R_1,R_2,R_3)$ is given by 

% \begin{widetext}
\begin{align}
e^{-\beta V_B^{(N=3)}} =  \frac{1}{6} \left( e^{-\beta E_{ooo}} +3e^{-\beta E_{Oo}} +2e^{-\beta E_{  \bigcirc }} \right) .
\label{eq:VB3_wdiagrams}  
\end{align}
% \end{widetext}
% Note the open one in Eq.~(\ref{eq:one}).

Here we define $E_{ooo} \equiv V_D^{(3)}$, $E_{Oo} \equiv E_o(R_1)+ E^{(2)}(R_2,R_3)$ and $E_{\bigcirc} \equiv E^{(3)}(R_1,R_2,R_3)$ and denote the total spring energy of a ring polymer constructed by connecting all of the beads of $k$ particles sequentially as

% \begin{widetext}
\begin{align}
% E_N^{(k)}(R_{N-k+1},...,R_N) \equiv \frac{1}{2}m \omega_P^2 \sum_{j=1}^{kP} \left( \bm r^{j+1} - \bm r^{j} \right) ^2.
E_N^{(k)}(R_{N-k+1},...,R_N) \equiv \frac{1}{2}m \omega_P^2 \sum_{l=N-k+1}^N \sum_{j=1}^P \left( \bm r_l^{j+1} - \bm r_l^{j} \right) ^2.
\label{eq:Ek}
\end{align}
% \end{widetext}
% Note the open one in Eq.~(\ref{eq:one}).
In Equation~\ref{eq:Ek} it is implied that $\bm r_N^{P+1} = \bm r_1^1$ or otherwise $\bm r_l^{P+1} = \bm r_{l+1}^1$. 
%Figure~\ref{fig:Fig1} presents the diagrammatic representation of the configurations contributing to Equation~\ref{eq:VB3_wdiagrams}. 
We note that $U(R_1,...,R_N)$ is identical to the case of distinguishable particles and that, using the Trotter decomposition, only the terms arising from the kinetic energy operator are affected by the permutations.
% In Equation~\ref{eq:Ek}, we have arranged the beads of particles $1,...,k$ to form a single vector $ \bm r = (\bm r_{N-k+1}^1,...,\bm r_{N-k+1}^P,...,\bm r_N^1,...,\bm r_N^P) $ of length $kP$ with the boundary conditions $ \bm r^{kP+1} = \bm r^1 $. 

\begin{figure}
\includegraphics[width=1.0\columnwidth]{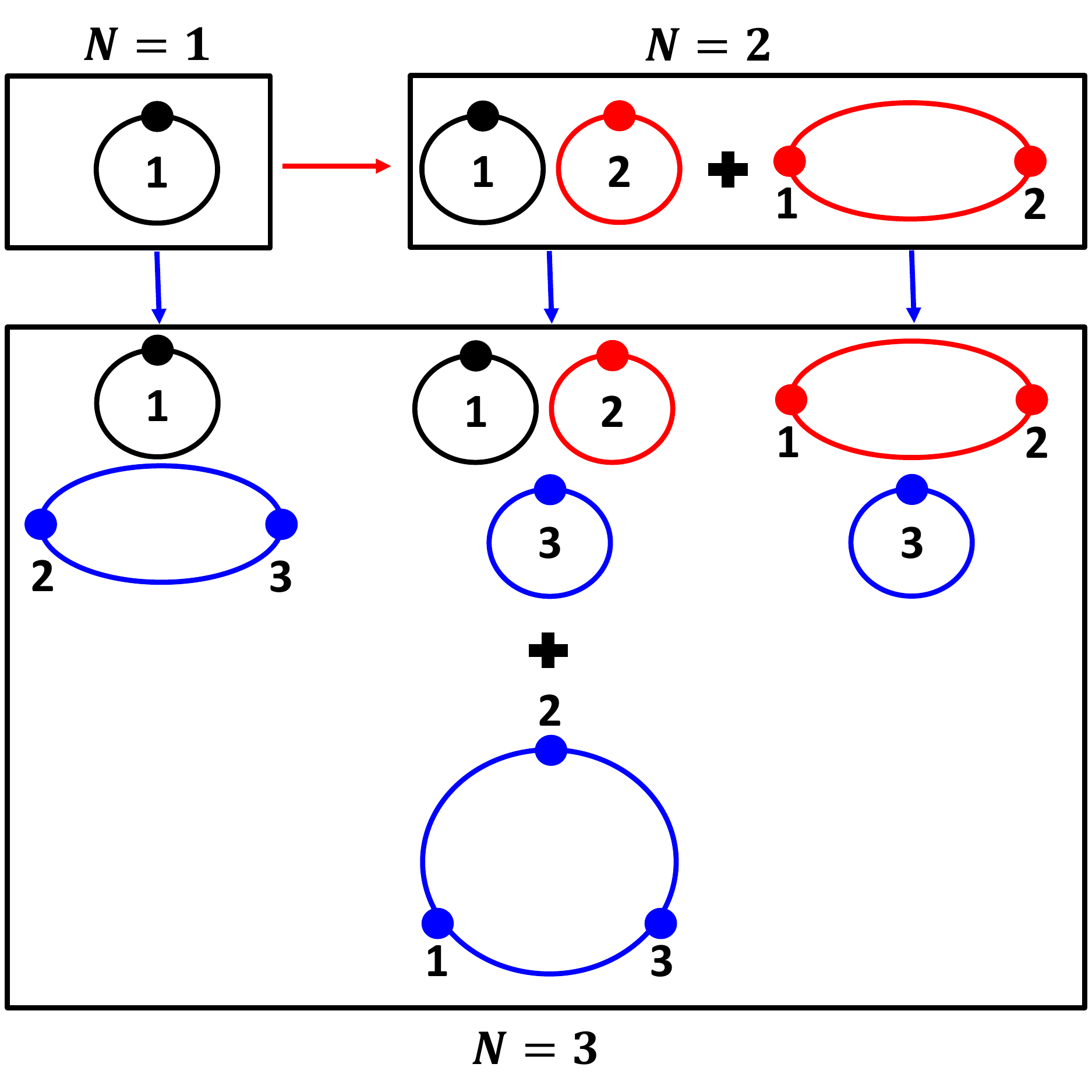}% Here is how to import EPS art
\caption{\label{fig:Fig2} Schematic drawing of the recursive procedure to generate all ring polymer configurations for $N=3$. See description in text.}
\end{figure}

While the computational cost is significantly reduced, as compared to summing over all permutations, the number of all the unique configurations still scales exponentially with system size. 
However, taking $N=3$ as an example, one finds that all configurations can be generated by the following procedure (see Figure~\ref{fig:Fig2}): Starting from a single ring polymer for particle 1, we generate all configurations for $N=2$ by adding a ring polymer representing particle 2, to the configuration of $N=1$ and then adding another configuration composed of particles $1-2$ connected sequentially to a long ring. We generate the configurations for $N=3$ by adding a ring of double-length, composed of particles $2-3$ connected sequentially, to the configuration of $N=1$ and a ring polymer representing particle 3 to the configurations of $N=2$. Lastly, we add another configuration composed of all 3 particles connected sequentially in a long ring. 

This procedure can be generalized to $N$ particles using the recurrence relation

\begin{equation}
e^{-\beta V_B^{(N)}} = 
\frac{1}{N} \sum_{k=1}^N e^{-\beta\left(E_N^{(k)}+V_B^{(N-k)}\right)},
\label{eq:VB_recursion}
\end{equation}

where $V_B^{(0)}=0$. The partition function is then given by

\begin{equation}
Z_B^{(N)} \sim \int e^{-\beta \left( V_B^{(N)} +U \right) } dR_1 ... dR_N .
\label{eq:ZB}
\end{equation}

The forces required for PIMD simulations, can also be written using a recurrence relation that is given in the SI. We note that in practical evaluation of Equation~\ref{eq:VB_recursion} it is useful to rewrite it so that only small energy differences appear in the exponent. Further details of the implementation are given in the SI.

We note that for non-interacting particles, the integration in Equation~\ref{eq:ZB} can be done analytically. Using  Equation~\ref{eq:VB_recursion}, this results in a known~\cite{Borrmann1993, Schmidt2002, krauth2006} recurrence relation for the partition function of $N$ non-interacting Bosons, $Z_B^{(N)}=\sum_{k=1}^N z_k Z_B^{(N-k)}$, where $z_k$ is the partition function of a single particle evaluated at inverse temperature $k\beta$.
%~\footnote{For non-interacting particles, $z_k$ is isomorphic to the partition function of $k$ particles connected sequentially to one ring at temperature $\beta$.}. 

A key observation is that Equation~\ref{eq:VB_recursion} can be used to evaluate the potential energy for $N$ Bosons, at a particular point $(R'_1,...,R'_N)$ without enumerating explicitly all ring-polymer configurations. This leads to an algorithm that scales cubically with system size (see SI). 
%We note in passing that Equation~\ref{eq:VB_recursion} for the potential is analogous to a recurrence relation for the cycle index polynomial of the symmetric group~\cite{Goldberg1993}.% in its natural action.[REF MISSING]

To evaluate thermal expectation values for quantum-mechanical operators using PIMD simulations, expressions for estimators must be derived. Estimators for operators which are local in coordinate representation are obtained as in PIMD simulations for distinguishable particles. For example, the estimator for the density is

\begin{equation}
\rho(\bm r)  = \left\langle  \frac{1}{P} \sum_{j=1}^P \sum_{l=1}^N  \delta (\bm r_l^j - \bm r)\right\rangle.
\label{eq:rho_estimator}
\end{equation}

For the energy, we derive
\begin{equation}
\left\langle E \right\rangle = \frac{PdN}{2\beta} + \left\langle U \right\rangle + \left\langle V_B^{(N)} +\beta \frac{\partial V_B^{(N)}} {\partial{\beta}} \right\rangle  ,
\label{eq:E_estimator}
\end{equation}

where $ V_B^{(N)} +\beta \frac{\partial V_B^{(N)}} {\partial{\beta}} $ is also calculated using a recurrence relation, given in the SI. It is well known that, for distinguishable particles, the variance of the thermodynamic estimator grows with the number of beads $P$~\cite{Herman1982}. Instead, the estimator above does is well-behaved due to the fact that all terms carry a positive sign. 

%\section{Results}

To test the method, we first applied it to systems of up to $N=64$ non-interacting Bosons. The agreement for the ground-state energy was excellent $(\sim 0.5\%)$ as shown in the SI. To verify that the correct Bose-Einstein statistics are obtained we also checked that the energy as a function of temperature agrees with analytical results (see SI).
To further validate the method, we performed simulations of small systems of interacting Bosons. For such small systems very accurate numerical results, obtained from diagonalizing the Hamiltonian, are available~\cite{Mujal2017}. Following Mujal \textit{et al.}, we simulate $N=2-4$ Bosons confined in an isotropic 2D harmonic trap and interacting through the repulsive Gaussian pair-potential:

\begin{equation}
V(|\bm r_l - \bm r_m|) = \frac{g}{\pi s^2}e^{-\frac{(\bm r_l - \bm r_m)^2}{s^2}}.
\label{eq:int_pot}
\end{equation}
In Equation~\ref{eq:int_pot}, $g$ and $s$ are a measure of the interaction strength and range, respectively. We take $s=0.5$ and we vary the interaction strength in the interval $g=0-16$ (in harmonic oscillator units). These values have been shown to span a wide range of interactions from weakly interacting particles, for which a mean-field approximation~\cite{Dalfovo1999} is valid, to strongly interacting particles, for which a correlated ansatz for the wave function must be used~\cite{Mujal2017}. Moreover, for small systems it was shown that the high values of $g$ lead to a pair correlation function which resembles that of a system of non-interacting Fermions of the same size~\cite{Mujal2018}.

\begin{figure}
\includegraphics[width=1\columnwidth]{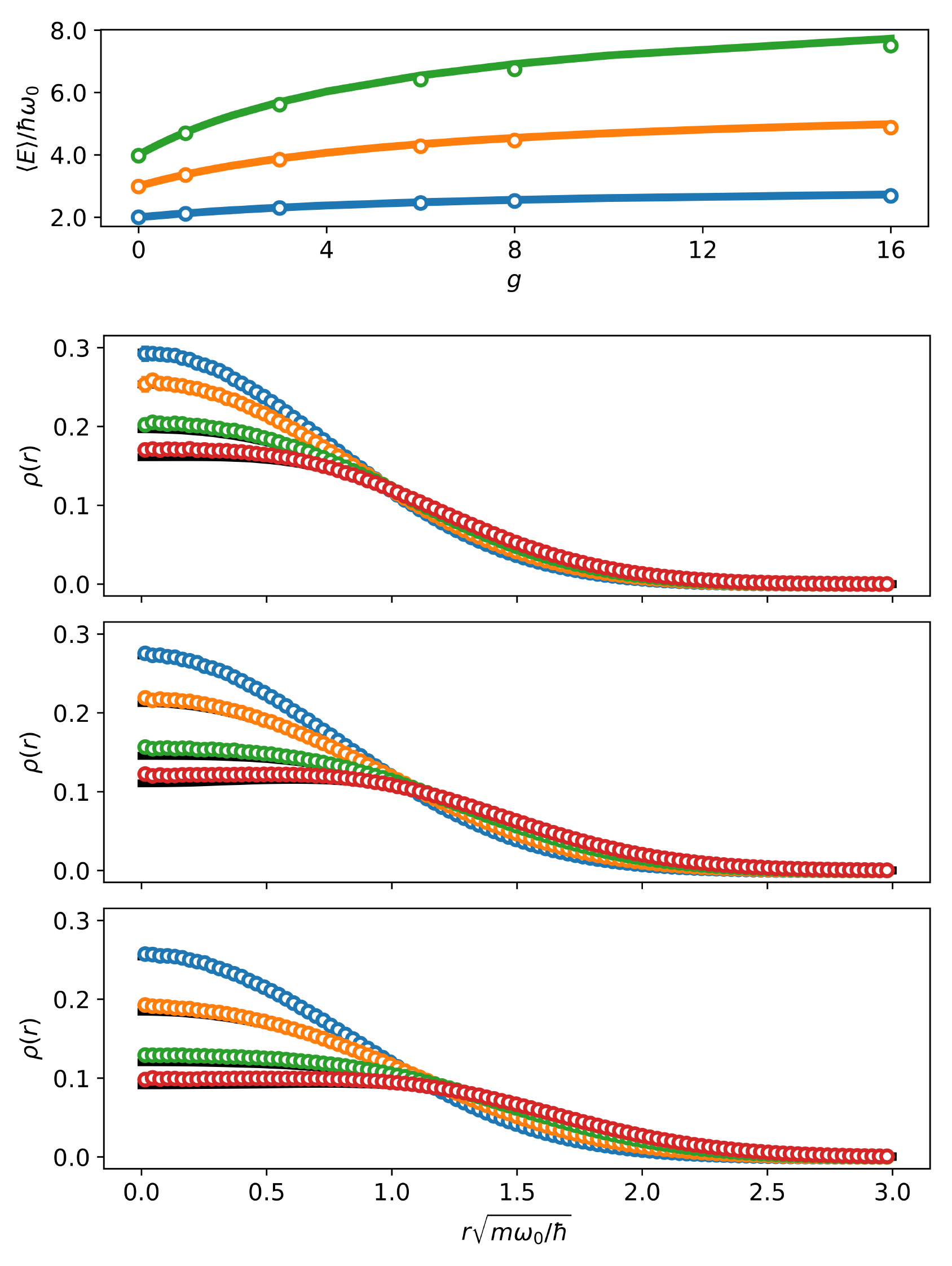}
\caption{\label{fig:Fig3} A comparison of the results of this Letter (circles) with those of ref.~\cite{Mujal2017} (solid lines). In the top panel, blue, orange and green symbols show the results for the total energy of $N=2,3$ and 4, respectively. The following panels, from top to bottom, show the density (normalized to unity) for $N=2,3$ and $4$, respectively. Blue, orange, green and red symbols represent $g=1,3,8$ and $16$, respectively.}
\end{figure}

The new method was implemented in a development version of LAMMPS~\cite{Plimpton1995}. For all the results below simulations using 72 beads were performed with a time step of 1 fs. Statistics was accumulated for 14 ns following an initial equilibration run of 1 ns. The statistical error was evaluated using block averaging. If not shown, it is smaller than the symbol size. A 2D trap frequency corresponding to $\hbar \omega_0 = 3$ meV was used. The simulations were performed with $m=1$ a.u. and at a temperature corresponding to $\beta \hbar \omega_0 = 6$ using the Nos\'{e}-Hoover chains thermostat~\cite{Martyna1992}.  The results were converged with respect to the number of beads at all range of interaction strengths (see SI for details). 

Figure~\ref{fig:Fig3} shows the energy as a function of $g$ for $N=2-4$. Very good agreement is obtained with the results of ref.~\cite{Mujal2017}. The maximum absolute deviation is ~2.9\% and the mean absolute deviation is 1.3\%. The comparison between the density obtained and exact diagonalization results is also shown. The agreement is very good and the mean absolute deviation for all range of interactions is $\sim 2.7\%$. This slight discrepancy could be due to the finite basis-set used in ref.~\cite{Mujal2017}. 

\begin{figure}
\includegraphics[width=1.0\columnwidth]{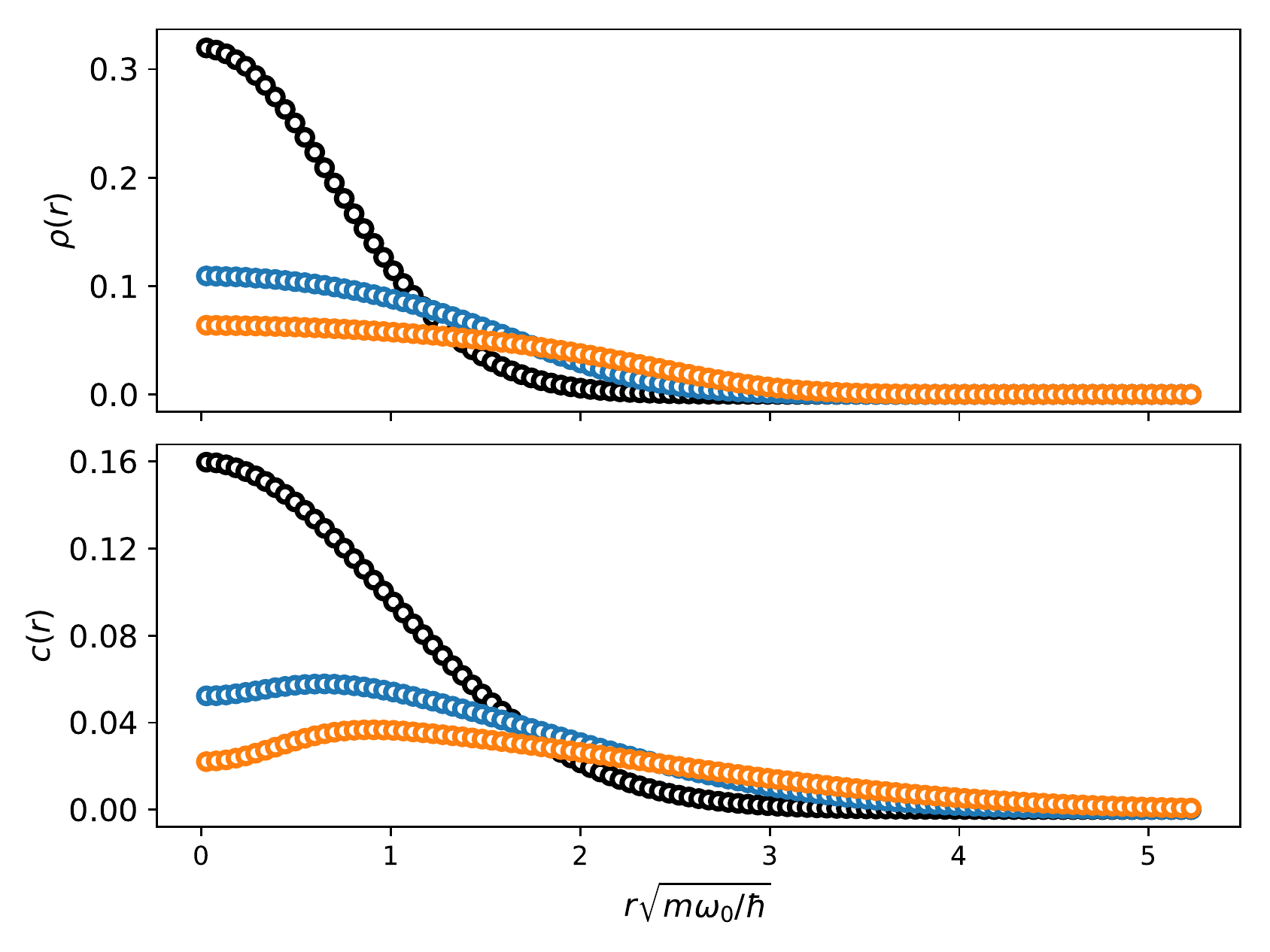}% Here is how to import EPS art
\caption{\label{fig:Fig4} The density (upper panel) and pair-correlation function (both normalized to unity) obtained for $N=32$ particles. Black, blue and orange circles represent $g=0,1$ and $3$, respectively. }
\end{figure}

%As mentioned before, the number of unique ring-polymer configurations grows exponentially with the number of particles $N$. Although $N=32$ might seem like a relatively-small system, already at this size, the number of unique ring-polymer configurations is 8349, which is prohibitive. 
%For example, for $N=32$ the number of configurations is 8349 while the number of energy terms required to evaluate the potential in equation~\ref{eq:VB_recursion} is only 528 (see SI for details). Therefore, already for systems with $N=32$ it is necessary to use the method presented in this paper. 
%We note also that for a periodic systems, finite-size effects are already small for this $N$~\cite{Boninsegni2006}.  

As mentioned previously, the number of ring-polymer configurations scales exponentially with $N$. For $N=32$ the number of unique configurations is 8439 while the number of energy terms required to evaluate Equation~\ref{eq:VB_recursion} is only 528 (see SI for details). Therefore, we chose $N=32$ to demonstrate the ability of the method to handle a large number of permutations. We used $g=3$ since already for such a repulsive potential significant differences from the non-interacting case are observed. The simulations were performed with $P=36$. Other simulation parameters are identical to the ones used for $N=2-4$.
Figure~\ref{fig:Fig4} presents the density and pair-correlation function. Due to the repulsion between the particles, the density at the center of the trap is significantly lowered and the probability to find particles away from the center of the trap is increased. The pair-correlation function shows a
complementary trend. The probability of finding any two particles at the same position is lowered as the repulsive interaction is introduced and the most probable distance between the two particles is $r\sqrt{m \omega_0 / \hbar } \approx 0.9$ for $g=3$. 

\begin{figure}
\includegraphics[width=1\columnwidth]{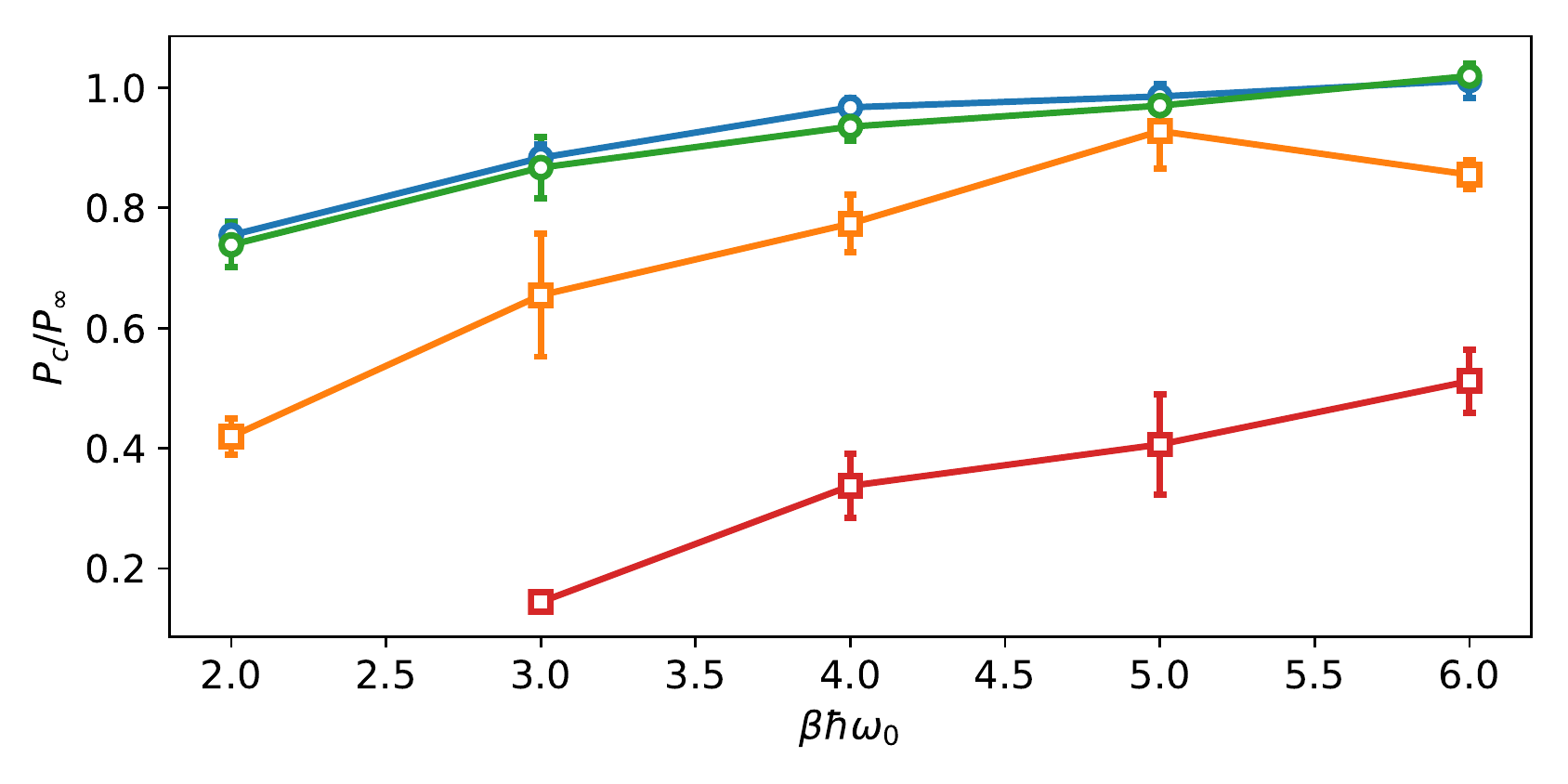}% Here is how to import EPS art
\caption{\label{fig:Fig5} The average probability for the longest ring-polymer configuration. Blue and green circles represent $N=2$ and $32$ non-interacting particles, respectively. Orange and red squares represent results for two particles using $g=8$ and 32 particles using $g=3$, respectively. $P_{\infty}$ is $1/N$.}
\end{figure}

For superfluid Helium, Ceperley~\cite{Ceperley1995} has shown that the probability of a particle to be incorporated into a long ring, composed of several particles, grows significantly below the critical temperature.  Therefore, the relative probability of ring-polymer configurations of different lengths can serve as an indication of the importance of exchange effects. Moreover, it was shown that the probability for the longest ring tends to $1/N$ at low temperatures. 
This analysis is readily performed in PIMC simulations, in which permutations are individually introduced. However, it is desirable also for PIMD simulations. To this effect, we define the instantaneous probability of a given ring-polymer configuration as 
\begin{align}
P_c = \frac{w_c \cdot e^{-\beta E_c}}{ \frac{1}{N} \sum_{k=1}^N e^{-\beta\left(E_N^{(k)}+V_B^{(N-k)}\right)} },
\label{eq:pEO}  
\end{align}
where $w_c$ is the number of permutations pertaining to configuration $c$.
The average probability for $N=2$ and $32$ is presented in Figure~\ref{fig:Fig5}. It can be seen that the probability for the ring-polymer configuration in which all $N$ particles are connected in a ring grows with decreasing temperature. Moreover, for $g=0$ it has the correct asymptotic behaviour already at $\beta \hbar \omega_0 =6 $. Introducing repulsive interactions, the increase in probability is slower, which suggests that exchange effects become more important at temperatures lower than those of a non-interacting system. The probability for the configuration representing distinguishable particles decreases with decreasing temperature (not shown). For $N=32$ it is negligible, compared to the longest ring-polymer configuration, at all temperatures studied.

%\section{Conclusions}

In conclusion, the aim of this paper is to present a new method for performing finite-temperature simulations of Bosons. The method has been tested and applied to systems of trapped Bosons with up to $64$  non-interacting and $32$ interacting particles. This represents an order of magnitude increase in system size, compared to diagonalization of the Hamiltonian~\cite{Mujal2017} and previous PIMD simulations of Bosons~\cite{Miura2000}. Our approach is not limited to systems of this size and we have already been experimenting with much larger systems. Since the computational cost scales relatively modestly with the number of particles, we hope that it will find useful applications in the study of cold atoms. For example, one could imagine using this approach in combination with ring-polymer MD~\cite{Habershon2013} to study Boson dynamics.

%\section{Acknowledgements}

\begin{acknowledgments}
We thank P. Mujal for sharing his raw data from ref.~\cite{Mujal2017} with us. BH would like to thank O. Buchman, J. Runeson and M. Nava for useful conversations and comments. The authors acknowledge support from the European Union Grant No. ERC-2014-ADG-670227/VARMET. Calculations were carried out on the Euler cluster at ETH Zurich. 
\end{acknowledgments}

\appendix

\bibliography{ms}% Produces the bibliography via BibTeX.

%merlin.mbs apsrev4-1.bst 2010-07-25 4.21a (PWD, AO, DPC) hacked
%Control: key (0)
%Control: author (8) initials jnrlst
%Control: editor formatted (1) identically to author
%Control: production of article title (-1) disabled
%Control: page (0) single
%Control: year (1) truncated
%Control: production of eprint (0) enabled
\begin{thebibliography}{34}%
\makeatletter
\providecommand \@ifxundefined [1]{%
 \@ifx{#1\undefined}
}%
\providecommand \@ifnum [1]{%
 \ifnum #1\expandafter \@firstoftwo
 \else \expandafter \@secondoftwo
 \fi
}%
\providecommand \@ifx [1]{%
 \ifx #1\expandafter \@firstoftwo
 \else \expandafter \@secondoftwo
 \fi
}%
\providecommand \natexlab [1]{#1}%
\providecommand \enquote  [1]{``#1''}%
\providecommand \bibnamefont  [1]{#1}%
\providecommand \bibfnamefont [1]{#1}%
\providecommand \citenamefont [1]{#1}%
\providecommand \href@noop [0]{\@secondoftwo}%
\providecommand \href [0]{\begingroup \@sanitize@url \@href}%
\providecommand \@href[1]{\@@startlink{#1}\@@href}%
\providecommand \@@href[1]{\endgroup#1\@@endlink}%
\providecommand \@sanitize@url [0]{\catcode `\\12\catcode `\$12\catcode
  `\&12\catcode `\#12\catcode `\^12\catcode `\_12\catcode `\%12\relax}%
\providecommand \@@startlink[1]{}%
\providecommand \@@endlink[0]{}%
\providecommand \url  [0]{\begingroup\@sanitize@url \@url }%
\providecommand \@url [1]{\endgroup\@href {#1}{\urlprefix }}%
\providecommand \urlprefix  [0]{URL }%
\providecommand \Eprint [0]{\href }%
\providecommand \doibase [0]{https://doi.org/}%
\providecommand \selectlanguage [0]{\@gobble}%
\providecommand \bibinfo  [0]{\@secondoftwo}%
\providecommand \bibfield  [0]{\@secondoftwo}%
\providecommand \translation [1]{[#1]}%
\providecommand \BibitemOpen [0]{}%
\providecommand \bibitemStop [0]{}%
\providecommand \bibitemNoStop [0]{.\EOS\space}%
\providecommand \EOS [0]{\spacefactor3000\relax}%
\providecommand \BibitemShut  [1]{\csname bibitem#1\endcsname}%
\let\auto@bib@innerbib\@empty
%</preamble>
\bibitem [{\citenamefont {Anderson}\ \emph {et~al.}(1995)\citenamefont
  {Anderson}, \citenamefont {Ensher}, \citenamefont {Matthews}, \citenamefont
  {Wieman},\ and\ \citenamefont {Cornell}}]{Anderson1995}%
  \BibitemOpen
  \bibfield  {author} {\bibinfo {author} {\bibfnamefont {M.~H.}\ \bibnamefont
  {Anderson}}, \bibinfo {author} {\bibfnamefont {J.~R.}\ \bibnamefont
  {Ensher}}, \bibinfo {author} {\bibfnamefont {M.~R.}\ \bibnamefont
  {Matthews}}, \bibinfo {author} {\bibfnamefont {C.~E.}\ \bibnamefont
  {Wieman}}, and\ \bibinfo {author} {\bibfnamefont {E.~A.}\ \bibnamefont
  {Cornell}},\ }\bibfield  {title} {\bibinfo {title} {{Observation of
  Bose-Einstein Condensation in a Dilute Atomic Vapor}},\ }\href
  {https://doi.org/10.1126/science.269.5221.198} {\bibfield  {journal}
  {\bibinfo  {journal} {Science}\ }\textbf {\bibinfo {volume} {269}},\ \bibinfo
  {pages} {198} (\bibinfo {year} {1995})}\BibitemShut {NoStop}%
\bibitem [{\citenamefont {Davis}\ \emph {et~al.}(1995)\citenamefont {Davis},
  \citenamefont {Mewes}, \citenamefont {Andrews}, \citenamefont {van Druten},
  \citenamefont {Durfee}, \citenamefont {Kurn},\ and\ \citenamefont
  {Ketterle}}]{Davis1995}%
  \BibitemOpen
  \bibfield  {author} {\bibinfo {author} {\bibfnamefont {K.~B.}\ \bibnamefont
  {Davis}}, \bibinfo {author} {\bibfnamefont {M.~O.}\ \bibnamefont {Mewes}},
  \bibinfo {author} {\bibfnamefont {M.~R.}\ \bibnamefont {Andrews}}, \bibinfo
  {author} {\bibfnamefont {N.~J.}\ \bibnamefont {van Druten}}, \bibinfo
  {author} {\bibfnamefont {D.~S.}\ \bibnamefont {Durfee}}, \bibinfo {author}
  {\bibfnamefont {D.~M.}\ \bibnamefont {Kurn}}, and\ \bibinfo {author}
  {\bibfnamefont {W.}~\bibnamefont {Ketterle}},\ }\bibfield  {title} {\bibinfo
  {title} {{Bose-Einstein Condensation in a Gas of Sodium Atoms}},\ }\href
  {https://doi.org/10.1103/PhysRevLett.75.3969} {\bibfield  {journal} {\bibinfo
   {journal} {Physical Review Letters}\ }\textbf {\bibinfo {volume} {75}},\
  \bibinfo {pages} {3969} (\bibinfo {year} {1995})}\BibitemShut {NoStop}%
\bibitem [{\citenamefont {Tanzi}\ \emph {et~al.}(2019)\citenamefont {Tanzi},
  \citenamefont {Lucioni}, \citenamefont {Fam{\`{a}}}, \citenamefont {Catani},
  \citenamefont {Fioretti}, \citenamefont {Gabbanini}, \citenamefont {Bisset},
  \citenamefont {Santos},\ and\ \citenamefont {Modugno}}]{Tanzi2019}%
  \BibitemOpen
  \bibfield  {author} {\bibinfo {author} {\bibfnamefont {L.}~\bibnamefont
  {Tanzi}}, \bibinfo {author} {\bibfnamefont {E.}~\bibnamefont {Lucioni}},
  \bibinfo {author} {\bibfnamefont {F.}~\bibnamefont {Fam{\`{a}}}}, \bibinfo
  {author} {\bibfnamefont {J.}~\bibnamefont {Catani}}, \bibinfo {author}
  {\bibfnamefont {A.}~\bibnamefont {Fioretti}}, \bibinfo {author}
  {\bibfnamefont {C.}~\bibnamefont {Gabbanini}}, \bibinfo {author}
  {\bibfnamefont {R.~N.}\ \bibnamefont {Bisset}}, \bibinfo {author}
  {\bibfnamefont {L.}~\bibnamefont {Santos}}, and\ \bibinfo {author}
  {\bibfnamefont {G.}~\bibnamefont {Modugno}},\ }\bibfield  {title} {\bibinfo
  {title} {{Observation of a Dipolar Quantum Gas with Metastable Supersolid
  Properties}},\ }\href {https://doi.org/10.1103/PhysRevLett.122.130405}
  {\bibfield  {journal} {\bibinfo  {journal} {Physical Review Letters}\
  }\textbf {\bibinfo {volume} {122}},\ \bibinfo {pages} {130405} (\bibinfo
  {year} {2019})}\BibitemShut {NoStop}%
\bibitem [{\citenamefont {Regal}\ \emph {et~al.}(2003)\citenamefont {Regal},
  \citenamefont {Ticknor}, \citenamefont {Bohn},\ and\ \citenamefont
  {Jin}}]{Regal2003}%
  \BibitemOpen
  \bibfield  {author} {\bibinfo {author} {\bibfnamefont {C.~A.}\ \bibnamefont
  {Regal}}, \bibinfo {author} {\bibfnamefont {C.}~\bibnamefont {Ticknor}},
  \bibinfo {author} {\bibfnamefont {J.~L.}\ \bibnamefont {Bohn}}, and\ \bibinfo
  {author} {\bibfnamefont {D.~S.}\ \bibnamefont {Jin}},\ }\bibfield  {title}
  {\bibinfo {title} {{Creation of ultracold molecules from a Fermi gas of
  atoms}},\ }\href {https://doi.org/10.1038/nature01738} {\bibfield  {journal}
  {\bibinfo  {journal} {Nature}\ }\textbf {\bibinfo {volume} {424}},\ \bibinfo
  {pages} {47} (\bibinfo {year} {2003})}\BibitemShut {NoStop}%
\bibitem [{\citenamefont {Ni}\ \emph {et~al.}(2008)\citenamefont {Ni},
  \citenamefont {Ospelkaus}, \citenamefont {de~Miranda}, \citenamefont {Pe'er},
  \citenamefont {Neyenhuis}, \citenamefont {Zirbel}, \citenamefont
  {Kotochigova}, \citenamefont {Julienne}, \citenamefont {Jin},\ and\
  \citenamefont {Ye}}]{Ni2008}%
  \BibitemOpen
  \bibfield  {author} {\bibinfo {author} {\bibfnamefont {K.-K.}\ \bibnamefont
  {Ni}}, \bibinfo {author} {\bibfnamefont {S.}~\bibnamefont {Ospelkaus}},
  \bibinfo {author} {\bibfnamefont {M.~H.~G.}\ \bibnamefont {de~Miranda}},
  \bibinfo {author} {\bibfnamefont {A.}~\bibnamefont {Pe'er}}, \bibinfo
  {author} {\bibfnamefont {B.}~\bibnamefont {Neyenhuis}}, \bibinfo {author}
  {\bibfnamefont {J.~J.}\ \bibnamefont {Zirbel}}, \bibinfo {author}
  {\bibfnamefont {S.}~\bibnamefont {Kotochigova}}, \bibinfo {author}
  {\bibfnamefont {P.~S.}\ \bibnamefont {Julienne}}, \bibinfo {author}
  {\bibfnamefont {D.~S.}\ \bibnamefont {Jin}}, and\ \bibinfo {author}
  {\bibfnamefont {J.}~\bibnamefont {Ye}},\ }\bibfield  {title} {\bibinfo
  {title} {{A High Phase-Space-Density Gas of Polar Molecules}},\ }\href
  {https://doi.org/10.1126/science.1163861} {\bibfield  {journal} {\bibinfo
  {journal} {Science}\ }\textbf {\bibinfo {volume} {322}},\ \bibinfo {pages}
  {231} (\bibinfo {year} {2008})}\BibitemShut {NoStop}%
\bibitem [{\citenamefont {Bakr}\ \emph {et~al.}(2010)\citenamefont {Bakr},
  \citenamefont {Peng}, \citenamefont {Tai}, \citenamefont {Ma}, \citenamefont
  {Simon}, \citenamefont {Gillen}, \citenamefont {Folling}, \citenamefont
  {Pollet},\ and\ \citenamefont {Greiner}}]{Bakr2010}%
  \BibitemOpen
  \bibfield  {author} {\bibinfo {author} {\bibfnamefont {W.~S.}\ \bibnamefont
  {Bakr}}, \bibinfo {author} {\bibfnamefont {A.}~\bibnamefont {Peng}}, \bibinfo
  {author} {\bibfnamefont {M.~E.}\ \bibnamefont {Tai}}, \bibinfo {author}
  {\bibfnamefont {R.}~\bibnamefont {Ma}}, \bibinfo {author} {\bibfnamefont
  {J.}~\bibnamefont {Simon}}, \bibinfo {author} {\bibfnamefont {J.~I.}\
  \bibnamefont {Gillen}}, \bibinfo {author} {\bibfnamefont {S.}~\bibnamefont
  {Folling}}, \bibinfo {author} {\bibfnamefont {L.}~\bibnamefont {Pollet}},
  and\ \bibinfo {author} {\bibfnamefont {M.}~\bibnamefont {Greiner}},\
  }\bibfield  {title} {\bibinfo {title} {{Probing the Superfluid-to-Mott
  Insulator Transition at the Single-Atom Level}},\ }\href
  {https://doi.org/10.1126/science.1192368} {\bibfield  {journal} {\bibinfo
  {journal} {Science}\ }\textbf {\bibinfo {volume} {329}},\ \bibinfo {pages}
  {547} (\bibinfo {year} {2010})}\BibitemShut {NoStop}%
\bibitem [{\citenamefont {Sherson}\ \emph {et~al.}(2010)\citenamefont
  {Sherson}, \citenamefont {Weitenberg}, \citenamefont {Endres}, \citenamefont
  {Cheneau}, \citenamefont {Bloch},\ and\ \citenamefont {Kuhr}}]{Sherson2010}%
  \BibitemOpen
  \bibfield  {author} {\bibinfo {author} {\bibfnamefont {J.~F.}\ \bibnamefont
  {Sherson}}, \bibinfo {author} {\bibfnamefont {C.}~\bibnamefont {Weitenberg}},
  \bibinfo {author} {\bibfnamefont {M.}~\bibnamefont {Endres}}, \bibinfo
  {author} {\bibfnamefont {M.}~\bibnamefont {Cheneau}}, \bibinfo {author}
  {\bibfnamefont {I.}~\bibnamefont {Bloch}}, and\ \bibinfo {author}
  {\bibfnamefont {S.}~\bibnamefont {Kuhr}},\ }\bibfield  {title} {\bibinfo
  {title} {{Single-atom-resolved fluorescence imaging of an atomic Mott
  insulator}},\ }\href {https://doi.org/10.1038/nature09378} {\bibfield
  {journal} {\bibinfo  {journal} {Nature}\ }\textbf {\bibinfo {volume} {467}},\
  \bibinfo {pages} {68} (\bibinfo {year} {2010})}\BibitemShut {NoStop}%
\bibitem [{\citenamefont {Paredes}\ \emph {et~al.}(2004)\citenamefont
  {Paredes}, \citenamefont {Widera}, \citenamefont {Murg}, \citenamefont
  {Mandel}, \citenamefont {F{\"{o}}lling}, \citenamefont {Cirac}, \citenamefont
  {Shlyapnikov}, \citenamefont {H{\"{a}}nsch},\ and\ \citenamefont
  {Bloch}}]{Paredes2004}%
  \BibitemOpen
  \bibfield  {author} {\bibinfo {author} {\bibfnamefont {B.}~\bibnamefont
  {Paredes}}, \bibinfo {author} {\bibfnamefont {A.}~\bibnamefont {Widera}},
  \bibinfo {author} {\bibfnamefont {V.}~\bibnamefont {Murg}}, \bibinfo {author}
  {\bibfnamefont {O.}~\bibnamefont {Mandel}}, \bibinfo {author} {\bibfnamefont
  {S.}~\bibnamefont {F{\"{o}}lling}}, \bibinfo {author} {\bibfnamefont
  {I.}~\bibnamefont {Cirac}}, \bibinfo {author} {\bibfnamefont {G.~V.}\
  \bibnamefont {Shlyapnikov}}, \bibinfo {author} {\bibfnamefont {T.~W.}\
  \bibnamefont {H{\"{a}}nsch}}, and\ \bibinfo {author} {\bibfnamefont
  {I.}~\bibnamefont {Bloch}},\ }\bibfield  {title} {\bibinfo {title}
  {{TonksâGirardeau gas of ultracold atoms in an optical lattice}},\
  }\href {https://doi.org/10.1038/nature02530} {\bibfield  {journal} {\bibinfo
  {journal} {Nature}\ }\textbf {\bibinfo {volume} {429}},\ \bibinfo {pages}
  {277} (\bibinfo {year} {2004})}\BibitemShut {NoStop}%
\bibitem [{\citenamefont {Kinoshita}\ \emph {et~al.}(2004)\citenamefont
  {Kinoshita}, \citenamefont {Wenger},\ and\ \citenamefont
  {Weiss}}]{Kinoshita2004}%
  \BibitemOpen
  \bibfield  {author} {\bibinfo {author} {\bibfnamefont {T.}~\bibnamefont
  {Kinoshita}}, \bibinfo {author} {\bibfnamefont {T.}~\bibnamefont {Wenger}},
  and\ \bibinfo {author} {\bibfnamefont {D.~S.}\ \bibnamefont {Weiss}},\
  }\bibfield  {title} {\bibinfo {title} {{Observation of a One-Dimensional
  Tonks-Girardeau Gas}},\ }\href {https://doi.org/10.1126/science.1100700}
  {\bibfield  {journal} {\bibinfo  {journal} {Science}\ }\textbf {\bibinfo
  {volume} {305}},\ \bibinfo {pages} {1125} (\bibinfo {year}
  {2004})}\BibitemShut {NoStop}%
\bibitem [{\citenamefont {Z{\"{u}}rn}\ \emph {et~al.}(2012)\citenamefont
  {Z{\"{u}}rn}, \citenamefont {Serwane}, \citenamefont {Lompe}, \citenamefont
  {Wenz}, \citenamefont {Ries}, \citenamefont {Bohn},\ and\ \citenamefont
  {Jochim}}]{Zurn2012}%
  \BibitemOpen
  \bibfield  {author} {\bibinfo {author} {\bibfnamefont {G.}~\bibnamefont
  {Z{\"{u}}rn}}, \bibinfo {author} {\bibfnamefont {F.}~\bibnamefont {Serwane}},
  \bibinfo {author} {\bibfnamefont {T.}~\bibnamefont {Lompe}}, \bibinfo
  {author} {\bibfnamefont {A.~N.}\ \bibnamefont {Wenz}}, \bibinfo {author}
  {\bibfnamefont {M.~G.}\ \bibnamefont {Ries}}, \bibinfo {author}
  {\bibfnamefont {J.~E.}\ \bibnamefont {Bohn}}, and\ \bibinfo {author}
  {\bibfnamefont {S.}~\bibnamefont {Jochim}},\ }\bibfield  {title} {\bibinfo
  {title} {{Fermionization of Two Distinguishable Fermions}},\ }\href
  {https://doi.org/10.1103/PhysRevLett.108.075303} {\bibfield  {journal}
  {\bibinfo  {journal} {Physical Review Letters}\ }\textbf {\bibinfo {volume}
  {108}},\ \bibinfo {pages} {075303} (\bibinfo {year} {2012})}\BibitemShut
  {NoStop}%
\bibitem [{\citenamefont {B{\"{u}}chler}\ \emph {et~al.}(2007)\citenamefont
  {B{\"{u}}chler}, \citenamefont {Micheli},\ and\ \citenamefont
  {Zoller}}]{Buchler2007}%
  \BibitemOpen
  \bibfield  {author} {\bibinfo {author} {\bibfnamefont {H.~P.}\ \bibnamefont
  {B{\"{u}}chler}}, \bibinfo {author} {\bibfnamefont {A.}~\bibnamefont
  {Micheli}}, and\ \bibinfo {author} {\bibfnamefont {P.}~\bibnamefont
  {Zoller}},\ }\bibfield  {title} {\bibinfo {title} {{Three-body interactions
  with cold polar molecules}},\ }\href {https://doi.org/10.1038/nphys678}
  {\bibfield  {journal} {\bibinfo  {journal} {Nature Physics}\ }\textbf
  {\bibinfo {volume} {3}},\ \bibinfo {pages} {726} (\bibinfo {year}
  {2007})}\BibitemShut {NoStop}%
\bibitem [{\citenamefont {Parrinello}\ and\ \citenamefont
  {Rahman}(1984)}]{Parrinello1984}%
  \BibitemOpen
  \bibfield  {author} {\bibinfo {author} {\bibfnamefont {M.}~\bibnamefont
  {Parrinello}}and\ \bibinfo {author} {\bibfnamefont {A.}~\bibnamefont
  {Rahman}},\ }\bibfield  {title} {\bibinfo {title} {{Study of an F center in
  molten KCl}},\ }\href {https://doi.org/10.1063/1.446740} {\bibfield
  {journal} {\bibinfo  {journal} {The Journal of Chemical Physics}\ }\textbf
  {\bibinfo {volume} {80}},\ \bibinfo {pages} {860} (\bibinfo {year}
  {1984})}\BibitemShut {NoStop}%
\bibitem [{\citenamefont {Feynman}\ and\ \citenamefont
  {Hibbs}(2005)}]{Feynman2005}%
  \BibitemOpen
  \bibfield  {author} {\bibinfo {author} {\bibfnamefont {R.~P.}\ \bibnamefont
  {Feynman}}and\ \bibinfo {author} {\bibfnamefont {A.~R.}\ \bibnamefont
  {Hibbs}},\ }\href@noop {} {\emph {\bibinfo {title} {{Quantum Mechanics and
  Path Integrals}}}}\ (\bibinfo  {publisher} {Dover pulications},\ \bibinfo
  {year} {2005})\BibitemShut {NoStop}%
\bibitem [{\citenamefont {Markland}\ and\ \citenamefont
  {Ceriotti}(2018)}]{Markland2018}%
  \BibitemOpen
  \bibfield  {author} {\bibinfo {author} {\bibfnamefont {T.~E.}\ \bibnamefont
  {Markland}}and\ \bibinfo {author} {\bibfnamefont {M.}~\bibnamefont
  {Ceriotti}},\ }\bibfield  {title} {\bibinfo {title} {{Nuclear quantum effects
  enter the mainstream}},\ }\href {https://doi.org/10.1038/s41570-017-0109}
  {\bibfield  {journal} {\bibinfo  {journal} {Nature Reviews Chemistry}\
  }\textbf {\bibinfo {volume} {2}},\ \bibinfo {pages} {0109} (\bibinfo {year}
  {2018})}\BibitemShut {NoStop}%
\bibitem [{\citenamefont {Chandler}\ and\ \citenamefont
  {Wolynes}(1981)}]{Chandler1981}%
  \BibitemOpen
  \bibfield  {author} {\bibinfo {author} {\bibfnamefont {D.}~\bibnamefont
  {Chandler}}and\ \bibinfo {author} {\bibfnamefont {P.~G.}\ \bibnamefont
  {Wolynes}},\ }\bibfield  {title} {\bibinfo {title} {{Exploiting the
  isomorphism between quantum theory and classical statistical mechanics of
  polyatomic fluids}},\ }\href {https://doi.org/10.1063/1.441588} {\bibfield
  {journal} {\bibinfo  {journal} {The Journal of Chemical Physics}\ }\textbf
  {\bibinfo {volume} {74}},\ \bibinfo {pages} {4078} (\bibinfo {year}
  {1981})}\BibitemShut {NoStop}%
\bibitem [{\citenamefont {Pollock}\ and\ \citenamefont
  {Ceperley}(1984)}]{Pollock1984}%
  \BibitemOpen
  \bibfield  {author} {\bibinfo {author} {\bibfnamefont {E.~L.}\ \bibnamefont
  {Pollock}}and\ \bibinfo {author} {\bibfnamefont {D.~M.}\ \bibnamefont
  {Ceperley}},\ }\bibfield  {title} {\bibinfo {title} {{Simulation of quantum
  many-body systems by path-integral methods}},\ }\href
  {https://doi.org/10.1103/PhysRevB.30.2555} {\bibfield  {journal} {\bibinfo
  {journal} {Physical Review B}\ }\textbf {\bibinfo {volume} {30}},\ \bibinfo
  {pages} {2555} (\bibinfo {year} {1984})}\BibitemShut {NoStop}%
\bibitem [{\citenamefont {Ceperley}(1995)}]{Ceperley1995}%
  \BibitemOpen
  \bibfield  {author} {\bibinfo {author} {\bibfnamefont {D.~M.}\ \bibnamefont
  {Ceperley}},\ }\bibfield  {title} {\bibinfo {title} {{Path integrals in the
  theory of condensed helium}},\ }\href
  {https://doi.org/10.1103/RevModPhys.67.279} {\bibfield  {journal} {\bibinfo
  {journal} {Reviews of Modern Physics}\ }\textbf {\bibinfo {volume} {67}},\
  \bibinfo {pages} {279} (\bibinfo {year} {1995})}\BibitemShut {NoStop}%
\bibitem [{\citenamefont {DuBois}\ \emph {et~al.}(2014)\citenamefont {DuBois},
  \citenamefont {Brown},\ and\ \citenamefont {Alder}}]{DuBois2014}%
  \BibitemOpen
  \bibfield  {author} {\bibinfo {author} {\bibfnamefont {J.~L.}\ \bibnamefont
  {DuBois}}, \bibinfo {author} {\bibfnamefont {E.~W.}\ \bibnamefont {Brown}},
  and\ \bibinfo {author} {\bibfnamefont {B.~J.}\ \bibnamefont {Alder}},\
  }\bibfield  {title} {\bibinfo {title} {{Overcoming the fermion sign problem
  in homogeneous systems}},\ }\href {http://arxiv.org/abs/1409.3262} {\
  (\bibinfo {year} {2014})},\ \Eprint {https://arxiv.org/abs/1409.3262}
  {arXiv:1409.3262} \BibitemShut {NoStop}%
\bibitem [{\citenamefont {Runeson}\ \emph {et~al.}(2018)\citenamefont
  {Runeson}, \citenamefont {Nava},\ and\ \citenamefont
  {Parrinello}}]{Runeson2018}%
  \BibitemOpen
  \bibfield  {author} {\bibinfo {author} {\bibfnamefont {J.}~\bibnamefont
  {Runeson}}, \bibinfo {author} {\bibfnamefont {M.}~\bibnamefont {Nava}}, and\
  \bibinfo {author} {\bibfnamefont {M.}~\bibnamefont {Parrinello}},\ }\bibfield
   {title} {\bibinfo {title} {{Quantum Symmetry from Enhanced Sampling
  Methods}},\ }\href {https://doi.org/10.1103/PhysRevLett.121.140602}
  {\bibfield  {journal} {\bibinfo  {journal} {Physical Review Letters}\
  }\textbf {\bibinfo {volume} {121}},\ \bibinfo {pages} {140602} (\bibinfo
  {year} {2018})}\BibitemShut {NoStop}%
\bibitem [{\citenamefont {Boninsegni}\ \emph {et~al.}(2006)\citenamefont
  {Boninsegni}, \citenamefont {Prokof'ev},\ and\ \citenamefont
  {Svistunov}}]{Boninsegni2006}%
  \BibitemOpen
  \bibfield  {author} {\bibinfo {author} {\bibfnamefont {M.}~\bibnamefont
  {Boninsegni}}, \bibinfo {author} {\bibfnamefont {N.}~\bibnamefont
  {Prokof'ev}}, and\ \bibinfo {author} {\bibfnamefont {B.}~\bibnamefont
  {Svistunov}},\ }\bibfield  {title} {\bibinfo {title} {{Worm Algorithm for
  Continuous-Space Path Integral Monte Carlo Simulations}},\ }\href
  {https://doi.org/10.1103/PhysRevLett.96.070601} {\bibfield  {journal}
  {\bibinfo  {journal} {Physical Review Letters}\ }\textbf {\bibinfo {volume}
  {96}},\ \bibinfo {pages} {070601} (\bibinfo {year} {2006})}\BibitemShut
  {NoStop}%
\bibitem [{\citenamefont {Miura}\ and\ \citenamefont
  {Okazaki}(2000)}]{Miura2000}%
  \BibitemOpen
  \bibfield  {author} {\bibinfo {author} {\bibfnamefont {S.}~\bibnamefont
  {Miura}}and\ \bibinfo {author} {\bibfnamefont {S.}~\bibnamefont {Okazaki}},\
  }\bibfield  {title} {\bibinfo {title} {{Path integral molecular dynamics for
  BoseâEinstein and FermiâDirac statistics}},\ }\href
  {https://doi.org/10.1063/1.481652} {\bibfield  {journal} {\bibinfo  {journal}
  {The Journal of Chemical Physics}\ }\textbf {\bibinfo {volume} {112}},\
  \bibinfo {pages} {10116} (\bibinfo {year} {2000})}\BibitemShut {NoStop}%
\bibitem [{\citenamefont {Mujal}\ \emph {et~al.}(2017)\citenamefont {Mujal},
  \citenamefont {Sarl{\'{e}}}, \citenamefont {Polls},\ and\ \citenamefont
  {Juli{\'{a}}-D{\'{i}}az}}]{Mujal2017}%
  \BibitemOpen
  \bibfield  {author} {\bibinfo {author} {\bibfnamefont {P.}~\bibnamefont
  {Mujal}}, \bibinfo {author} {\bibfnamefont {E.}~\bibnamefont {Sarl{\'{e}}}},
  \bibinfo {author} {\bibfnamefont {A.}~\bibnamefont {Polls}}, and\ \bibinfo
  {author} {\bibfnamefont {B.}~\bibnamefont {Juli{\'{a}}-D{\'{i}}az}},\
  }\bibfield  {title} {\bibinfo {title} {{Quantum correlations and degeneracy
  of identical bosons in a two-dimensional harmonic trap}},\ }\href
  {https://doi.org/10.1103/PhysRevA.96.043614} {\bibfield  {journal} {\bibinfo
  {journal} {Physical Review A}\ }\textbf {\bibinfo {volume} {96}},\ \bibinfo
  {pages} {043614} (\bibinfo {year} {2017})}\BibitemShut {NoStop}%
\bibitem [{\citenamefont {Tuckerman}(2010)}]{tuckerman2010}%
  \BibitemOpen
  \bibfield  {author} {\bibinfo {author} {\bibfnamefont {M.}~\bibnamefont
  {Tuckerman}},\ }\href@noop {} {\emph {\bibinfo {title} {{Statistical
  mechanics: theory and molecular simulation}}}}\ (\bibinfo  {publisher}
  {Oxford university press},\ \bibinfo {year} {2010})\BibitemShut {NoStop}%
\bibitem [{\citenamefont {Lyubartsev}\ and\ \citenamefont
  {Vorontsov-Velyaminov}(1993)}]{Lyubartsev1993}%
  \BibitemOpen
  \bibfield  {author} {\bibinfo {author} {\bibfnamefont {A.~P.}\ \bibnamefont
  {Lyubartsev}}and\ \bibinfo {author} {\bibfnamefont {P.~N.}\ \bibnamefont
  {Vorontsov-Velyaminov}},\ }\bibfield  {title} {\bibinfo {title}
  {{Path-integral Monte Carlo method in quantum statistics for a system of N
  identical fermions}},\ }\href {https://doi.org/10.1103/PhysRevA.48.4075}
  {\bibfield  {journal} {\bibinfo  {journal} {Physical Review A}\ }\textbf
  {\bibinfo {volume} {48}},\ \bibinfo {pages} {4075} (\bibinfo {year}
  {1993})}\BibitemShut {NoStop}%
\bibitem [{\citenamefont {Voznesenskiy}\ \emph {et~al.}(2009)\citenamefont
  {Voznesenskiy}, \citenamefont {Vorontsov-Velyaminov},\ and\ \citenamefont
  {Lyubartsev}}]{Voznesenskiy2009}%
  \BibitemOpen
  \bibfield  {author} {\bibinfo {author} {\bibfnamefont {M.~A.}\ \bibnamefont
  {Voznesenskiy}}, \bibinfo {author} {\bibfnamefont {P.~N.}\ \bibnamefont
  {Vorontsov-Velyaminov}}, and\ \bibinfo {author} {\bibfnamefont {A.~P.}\
  \bibnamefont {Lyubartsev}},\ }\bibfield  {title} {\bibinfo {title}
  {{Path-integralâexpanded-ensemble Monte Carlo method in treatment of the
  sign problem for fermions}},\ }\href
  {https://doi.org/10.1103/PhysRevE.80.066702} {\bibfield  {journal} {\bibinfo
  {journal} {Physical Review E}\ }\textbf {\bibinfo {volume} {80}},\ \bibinfo
  {pages} {066702} (\bibinfo {year} {2009})}\BibitemShut {NoStop}%
\bibitem [{\citenamefont {Borrmann}\ and\ \citenamefont
  {Franke}(1993)}]{Borrmann1993}%
  \BibitemOpen
  \bibfield  {author} {\bibinfo {author} {\bibfnamefont {P.}~\bibnamefont
  {Borrmann}}and\ \bibinfo {author} {\bibfnamefont {G.}~\bibnamefont
  {Franke}},\ }\bibfield  {title} {\bibinfo {title} {{Recursion formulas for
  quantum statistical partition functions}},\ }\href
  {https://doi.org/10.1063/1.464180} {\bibfield  {journal} {\bibinfo  {journal}
  {The Journal of Chemical Physics}\ }\textbf {\bibinfo {volume} {98}},\
  \bibinfo {pages} {2484} (\bibinfo {year} {1993})}\BibitemShut {NoStop}%
\bibitem [{\citenamefont {Schmidt}\ and\ \citenamefont
  {Schnack}(2002)}]{Schmidt2002}%
  \BibitemOpen
  \bibfield  {author} {\bibinfo {author} {\bibfnamefont {H.-J.}\ \bibnamefont
  {Schmidt}}and\ \bibinfo {author} {\bibfnamefont {J.}~\bibnamefont
  {Schnack}},\ }\bibfield  {title} {\bibinfo {title} {{Partition functions and
  symmetric polynomials}},\ }\href {https://doi.org/10.1119/1.1412643}
  {\bibfield  {journal} {\bibinfo  {journal} {American Journal of Physics}\
  }\textbf {\bibinfo {volume} {70}},\ \bibinfo {pages} {53} (\bibinfo {year}
  {2002})}\BibitemShut {NoStop}%
\bibitem [{\citenamefont {Krauth}(2006)}]{krauth2006}%
  \BibitemOpen
  \bibfield  {author} {\bibinfo {author} {\bibfnamefont {W.}~\bibnamefont
  {Krauth}},\ }\href@noop {} {\emph {\bibinfo {title} {{Statistical mechanics:
  algorithms and computations}}}},\ Vol.~\bibinfo {volume} {13}\ (\bibinfo
  {publisher} {OUP Oxford},\ \bibinfo {year} {2006})\BibitemShut {NoStop}%
\bibitem [{\citenamefont {Herman}\ \emph {et~al.}(1982)\citenamefont {Herman},
  \citenamefont {Bruskin},\ and\ \citenamefont {Berne}}]{Herman1982}%
  \BibitemOpen
  \bibfield  {author} {\bibinfo {author} {\bibfnamefont {M.~F.}\ \bibnamefont
  {Herman}}, \bibinfo {author} {\bibfnamefont {E.~J.}\ \bibnamefont {Bruskin}},
  and\ \bibinfo {author} {\bibfnamefont {B.~J.}\ \bibnamefont {Berne}},\
  }\bibfield  {title} {\bibinfo {title} {{On path integral Monte Carlo
  simulations}},\ }\href {https://doi.org/10.1063/1.442815} {\bibfield
  {journal} {\bibinfo  {journal} {The Journal of Chemical Physics}\ }\textbf
  {\bibinfo {volume} {76}},\ \bibinfo {pages} {5150} (\bibinfo {year}
  {1982})}\BibitemShut {NoStop}%
\bibitem [{\citenamefont {Dalfovo}\ \emph {et~al.}(1999)\citenamefont
  {Dalfovo}, \citenamefont {Giorgini}, \citenamefont {Pitaevskii},\ and\
  \citenamefont {Stringari}}]{Dalfovo1999}%
  \BibitemOpen
  \bibfield  {author} {\bibinfo {author} {\bibfnamefont {F.}~\bibnamefont
  {Dalfovo}}, \bibinfo {author} {\bibfnamefont {S.}~\bibnamefont {Giorgini}},
  \bibinfo {author} {\bibfnamefont {L.~P.}\ \bibnamefont {Pitaevskii}}, and\
  \bibinfo {author} {\bibfnamefont {S.}~\bibnamefont {Stringari}},\ }\bibfield
  {title} {\bibinfo {title} {{Theory of Bose-Einstein condensation in trapped
  gases}},\ }\href {https://doi.org/10.1103/RevModPhys.71.463} {\bibfield
  {journal} {\bibinfo  {journal} {Reviews of Modern Physics}\ }\textbf
  {\bibinfo {volume} {71}},\ \bibinfo {pages} {463} (\bibinfo {year}
  {1999})}\BibitemShut {NoStop}%
\bibitem [{\citenamefont {Mujal}\ \emph {et~al.}(2018)\citenamefont {Mujal},
  \citenamefont {Polls},\ and\ \citenamefont
  {Juli{\'{a}}-D{\'{i}}az}}]{Mujal2018}%
  \BibitemOpen
  \bibfield  {author} {\bibinfo {author} {\bibfnamefont {P.}~\bibnamefont
  {Mujal}}, \bibinfo {author} {\bibfnamefont {A.}~\bibnamefont {Polls}}, and\
  \bibinfo {author} {\bibfnamefont {B.}~\bibnamefont
  {Juli{\'{a}}-D{\'{i}}az}},\ }\bibfield  {title} {\bibinfo {title} {{Fermionic
  Properties of Two Interacting Bosons in a Two-Dimensional Harmonic Trap}},\
  }\href {https://doi.org/10.3390/condmat3010009} {\bibfield  {journal}
  {\bibinfo  {journal} {Condensed Matter}\ }\textbf {\bibinfo {volume} {3}},\
  \bibinfo {pages} {9} (\bibinfo {year} {2018})}\BibitemShut {NoStop}%
\bibitem [{\citenamefont {Plimpton}(1995)}]{Plimpton1995}%
  \BibitemOpen
  \bibfield  {author} {\bibinfo {author} {\bibfnamefont {S.}~\bibnamefont
  {Plimpton}},\ }\bibfield  {title} {\bibinfo {title} {{Fast Parallel
  Algorithms for Short-Range Molecular Dynamics}},\ }\href
  {https://doi.org/10.1006/jcph.1995.1039} {\bibfield  {journal} {\bibinfo
  {journal} {Journal of Computational Physics}\ }\textbf {\bibinfo {volume}
  {117}},\ \bibinfo {pages} {1} (\bibinfo {year} {1995})}\BibitemShut {NoStop}%
\bibitem [{\citenamefont {Martyna}\ \emph {et~al.}(1992)\citenamefont
  {Martyna}, \citenamefont {Klein},\ and\ \citenamefont
  {Tuckerman}}]{Martyna1992}%
  \BibitemOpen
  \bibfield  {author} {\bibinfo {author} {\bibfnamefont {G.~J.}\ \bibnamefont
  {Martyna}}, \bibinfo {author} {\bibfnamefont {M.~L.}\ \bibnamefont {Klein}},
  and\ \bibinfo {author} {\bibfnamefont {M.}~\bibnamefont {Tuckerman}},\
  }\bibfield  {title} {\bibinfo {title} {{Nos{\'{e}}âHoover chains: The
  canonical ensemble via continuous dynamics}},\ }\href
  {https://doi.org/10.1063/1.463940} {\bibfield  {journal} {\bibinfo  {journal}
  {The Journal of Chemical Physics}\ }\textbf {\bibinfo {volume} {97}},\
  \bibinfo {pages} {2635} (\bibinfo {year} {1992})}\BibitemShut {NoStop}%
\bibitem [{\citenamefont {Habershon}\ \emph {et~al.}(2013)\citenamefont
  {Habershon}, \citenamefont {Manolopoulos}, \citenamefont {Markland},\ and\
  \citenamefont {Miller}}]{Habershon2013}%
  \BibitemOpen
  \bibfield  {author} {\bibinfo {author} {\bibfnamefont {S.}~\bibnamefont
  {Habershon}}, \bibinfo {author} {\bibfnamefont {D.~E.}\ \bibnamefont
  {Manolopoulos}}, \bibinfo {author} {\bibfnamefont {T.~E.}\ \bibnamefont
  {Markland}}, and\ \bibinfo {author} {\bibfnamefont {T.~F.}\ \bibnamefont
  {Miller}},\ }\bibfield  {title} {\bibinfo {title} {{Ring-Polymer Molecular
  Dynamics: Quantum Effects in Chemical Dynamics from Classical Trajectories in
  an Extended Phase Space}},\ }\href
  {https://doi.org/10.1146/annurev-physchem-040412-110122} {\bibfield
  {journal} {\bibinfo  {journal} {Annual Review of Physical Chemistry}\
  }\textbf {\bibinfo {volume} {64}},\ \bibinfo {pages} {387} (\bibinfo {year}
  {2013})}\BibitemShut {NoStop}%
\end{thebibliography}%


%apsrev4-2.bst 2018-12-27 (MD) hand-edited version of apsrev4-1.bst
%Control: key (0)
%Control: author (8) initials jnrlst
%Control: editor formatted (1) identically to author
%Control: production of article title (0) allowed
%Control: page (0) single
%Control: year (1) truncated
%Control: production of eprint (0) enabled
%

\end{document}